\pdfoutput=1

\def\editmode{0}
\def\reportmode{0}

\def\bibfilenames{WISENET}

\if\reportmode1
\documentclass[10pt,final,onecolumn]{IEEEtran}
\else
\documentclass{article}
\usepackage{spconf}
\fi

\if\editmode1  %%%%%%%%%%%%%%%%%%%%%%%%%%%%%%%%%%%%%%%%%%%%%%%%%%%%%%%%%%%
\usepackage[backend=bibtex,style=alphabetic,sorting=debug]{biblatex}
\DeclareFieldFormat{labelalpha}{\thefield{entrykey}}
\DeclareFieldFormat{extraalpha}{}
\bibliography{\bibfilenames}
\newcommand{\cmt}[1]{\noindent\textcolor{lightgreen}{\underline{[#1]}}} % comment
\newcommand{\hc}[1]{\textcolor{blue}{#1}} % highlight command --> to
\newenvironment{myitemize}{\begin{itemize}}{\end{itemize}}
\newcommand{\myitem}{\item}

\else
\usepackage{cite}
\newcommand{\cmt}[1]{} % comment
\newcommand{\hc}[1]{\textcolor{black}{#1}} % highlight command -->
\newenvironment{myitemize}{}{}
\newcommand{\myitem}{}

\fi %%%%%%%%%%%%%%%%%%%%%%%%%%%%%%%%%%%%%%%%%%%%%%%%%%%%%%%%%%%%%%%%%%%%%%

\usepackage{fixltx2e}

\usepackage{graphicx}

\usepackage{subcaption}              %subfigures with \mbox and \subfigure[]

\usepackage[utf8]{inputenc}

\usepackage{amsfonts}
\usepackage{amsmath}
\usepackage{mathtools}
\usepackage{amssymb}

\usepackage{bm}

\usepackage{color,verbatim}
\usepackage{multirow}
\usepackage{accents}

\usepackage{theoremref}
\usepackage{url}

\newcounter{rulecounter}
\newcommand{\resetrule}{ \setcounter{rulecounter}{0}}
\resetrule

\newsavebox{\selvestebox}
\newenvironment{colbox}[1]
  {\newcommand\colboxcolor{#1}%
   \begin{lrbox}{\selvestebox}%
   \begin{minipage}{\dimexpr\columnwidth-2\fboxsep\relax}}
  {\end{minipage}\end{lrbox}%
   \begin{center}
   \colorbox{\colboxcolor}{\usebox{\selvestebox}}
   \end{center}}

\definecolor{orange}{rgb}{1,0.8,0}
\definecolor{gray}{rgb}{.9,0.9,0.9}
\definecolor{darkgray}{rgb}{.3,0.3,0.3}
\definecolor{darkblue}{rgb}{.1,0.0,0.3}
\definecolor{lightblue}{rgb}{0.7,0.7,1}
\definecolor{lightred}{rgb}{1,0.7,.7}
\definecolor{purple}{RGB}{204,153,255}
\definecolor{lightgray}{rgb}{.95,0.95,0.95}
\definecolor{lightgreen}{rgb}{0.3,0.5,0.3}
\definecolor{darkgreen}{rgb}{0.05,0.3,0.05}

\newcommand{\ra}{$\rightarrow$~}

\newtheorem{myproposition}{Proposition}
\newtheorem{myremark}{Remark}
\newtheorem{myproblemstatement}{Problem Statement}
\newtheorem{mylemma}{Lemma}
\newtheorem{mytheorem}{Theorem}
\newtheorem{mydefinition}{Definition}
\newtheorem{mycorollary}{Corollary}

\usepackage{ mathrsfs }

\usepackage{algorithm}
\usepackage[noend]{algpseudocode}
\floatname{algorithm}{Procedure}

\begin{document}

\newcommand{\acom}[1]{\textcolor{red}{\textbf{[#1]}}} % comment
%\renewcommand{\acom}[1]{}

%%%%%%%%%%%%%%%%%%%%%%%%%%%%%%%%%%%%%%%%%%%%%%%%%%%%%%%%%%%%%%%%%%%%%
\title{Dynamic Network Identification From \\Non-stationary Vector Autoregressive Time Series}
%%%%%%%%%%%%%%%%%%%%%%%%%%%%%%%%%%%%%%%%%%%%%%%%%%%%%%%%%%%%%%%%%%%%%

\if\reportmode1
  \author{Luis M. Lopez-Ramos, Daniel Romero, Bakht Zaman, and Baltasar Beferull-Lozano\\ \today}
\else
\name{Luis M. Lopez-Ramos, %~\IEEEmembership{Member,~IEEE,}
Daniel Romero, %~\IEEEmembership{Member,~IEEE,}\\
  Bakht Zaman, %~\IEEEmembership{Student Member,~IEEE}, 
  and Baltasar Beferull-Lozano%~\IEEEmembership{Senior Member,~IEEE}}
 \thanks{The work in this paper was supported by the SFI Offshore Mechatronics grant 237896/O30.}}
\address{Intelligent Signal Processing and Wireless Networks Laboratory (WISENET) \\
Department of ICT, University of Agder, Grimstad, Norway \\
Email: \{luismiguel.lopez, daniel.romero, bakht.zaman, baltasar.beferull\}@uia.no}
\fi

\maketitle
\begin{abstract}
Learning the dynamics of complex systems features a large number of
applications in data science. 
%Graph-based tools constitute an appealing approach to accomplish such a task 
Graph-based modeling and inference underpins the most prominent
family of approaches to learn complex dynamics due to their ability to
capture the intrinsic sparsity of direct interactions in such systems.
They also provide the user with interpretable graphs that unveil
behavioral patterns and changes. To cope with the time-varying nature of
interactions, this paper develops an estimation criterion and a solver
to learn the parameters of a time-varying vector autoregressive model
supported on a network of time series. The notion of local breakpoint
is proposed to accommodate changes at individual edges. It contrasts
with existing works, which assume that changes at all nodes are aligned in
time. Numerical experiments validate the proposed schemes.
\end{abstract}

\if\reportmode0
\fi

\section{Introduction}
\label{sec:intro}

\cmt{overview}
\begin{myitemize}
\myitem\cmt{complex systems}Understanding the interactions among the
parts of a complex dynamic system lies at the core of data science
itself and countless applications in biology, sociology, neuroscience,
finance, as well as engineering realms such as cybernetics, mechatronics, and
control of industrial processes.
\myitem\cmt{causality}Successfully learning the presence or evolution
of these interactions allows forecasting and unveils complex behaviors
typically by spotting causality relations~\cite{granger1988causality}. 
\myitem\cmt{data driven\ra nets}To cope with the ever increasing  complexity
of the analyzed systems, traditional model-based paradigms are giving
way to the more contemporary data-driven perspectives,
where network-based approaches enjoy
great popularity due to their ability to both discern between direct
and indirect causality relations as well as to provide interaction
graphs amenable to intuitive human interpretation.
\myitem\cmt{non-stationary}In this context, the  time-varying nature 
of these interactions motivates inference schemes capable of handling
non-stationarity multivariate data.
\end{myitemize}

%% \begin{myitemize}
%% \myitem %Because of the presence of actuators and changes in environment conditions, 
%% The relationships between the sensor variables: \begin{myitemize}
%% \myitem have a non-stationary nature and change parsimoniously over time;
%% \myitem have memory, meaning that the signals depend on their own recent past;
%% \myitem can often be globally explained as a superposition of a small number of local interactions (network effect).
%% \end{myitemize}
%% %\myitem The evolution of the aforementioned spatio-temporal (causal) relationships (influences) can be assumed parsimonious, so that the relationships at a given time instant depend also on the relationships that exist at previous time instants. 
%% \cmt{motivation}
%% \myitem When dealing with time-varying models, detecting the so-called \emph{structural break points}, (instants where model parameters change significantly) and consistent parameter estimation are important, because \begin{myitemize}
%%     \myitem they may reveal important changes in the system underlying the observed process%, often providing important scientific insights.
%%     \myitem %Also, the sequence edge sets of the inferred time-varying graph 
%%     they inform about how causal relations (in the Granger \cite{granger1988causality} sense) evolve over time.
%% \end{myitemize}
%% \myitem In order to successfully recover and analyze models, the algorithms need to incorporate assumptions on how the model evolves over time and space.
%% \end{myitemize}
%% %relationships between the sensor variables that determine the behaviour of the process. 

\cmt{related work}\begin{myitemize}
%%
% VAR+non-stationary, NO graph structure
%%
\myitem\cmt{VAR+non-stationary, NO graph structure}Inference from
multiple time series has been traditionally addressed through vector
autoregressive (VAR) models~\cite{lutkepohl2005}.
\begin{myitemize}\myitem\cmt{categories}To cope with non-stationarity,  VAR
coefficients are assumed to
\begin{myitemize} 
\myitem\cmt{coefficients change smoothly}evolve smoothly over 
time~\cite{sato2007wavelet, dahlhaus2012locally, 
niedzwiecki2017adaptive},
\myitem\cmt{coefficients change according to a Markov chain}
to vary according to a hidden Markov
model~\cite{fox2008nonparametric}, 
\myitem\cmt{piecewise constant}or to remain constant over time
intervals separated by  \emph{structural breakpoints}~
\cite{ombao2005slex,aue2009break,cho2015multiple,cho2016change,safikhani2017structuralbreak,tank2017efficient,ohlsson2010segmentation}.
%% %
%% see~\cite{davis2006structural,
%% chan2014group} for univariate alternatives. 
\end{myitemize}\myitem\cmt{limitations}Due to the high number of effective degrees of
freedom of their models, these schemes can only satisfactorily
estimate VAR coefficients if the data generating system experiences slow
changes over time. 
\myitem\cmt{netw. structure necessary}To alleviate
this difficulty, a natural approach is to exploit the fact that
interactions among different parts of a complex system are
generally \emph{mediated}. For example, in an industrial plant where
tank A is connected to B, B is connected to C, and C is not connected
to A, the pressure of a fluid in a tank A affects \emph{directly} the
pressure of tank B and \emph{indirectly} (through B) the pressure at
tank C.
% underlies the time series generation, which leaves certain structure unexploited.
\end{myitemize}
%\myitem In \cite{kuznetsov2015learning}, learning bounds for forecasting of non-stationary time series are introduced. Specifically, a discrepancy measure that can be estimated from the data is proposed and two forecasting algorithms are designed. The calculation of the discrepancy requires solving a non-convex problem.
%%
% Dynamic network, NO memory
%%
\myitem \cmt{Dynamic network, NO memory}Thus, a number of works
focused on non-stationary data introduce graphs to capture this notion
of \emph{direct} interactions, either relying on 
\begin{myitemize}\myitem\cmt{categories}\begin{myitemize}\myitem\cmt{graphical models}graphical models~\cite{Kolar2010Estimating,friedman2008graphicallasso,angelosante2011graphical,hallac2017network}
\myitem\cmt{structural equation models}or structural equation models~\cite{shen2017tensor,baingana2017tracking}.
\end{myitemize}\myitem\cmt{limitations}Unfortunately, these approaches can only deal with
memoryless interactions, which limits their applicability to many
real-world scenarios. 
\end{myitemize}\end{myitemize}%
\begin{myitemize}%
%%
% Memory and network structure, NOT dynamic
%%
\myitem\cmt{Memory and network structure, NOT dynamic}Schemes that do
account for memory and graph structure include
models \begin{myitemize}%
\myitem\cmt{categories}%
\begin{myitemize}%
\myitem\cmt{VAR + Networks} based on
VAR~\cite{bolstad2011groupsparse} and models based on 
%mei2016causal
\myitem\cmt{structural VAR}structural VAR models;
see~\cite{shen2016nonlinear} and references therein. 
\end{myitemize}%
\myitem\cmt{limitations}However, these methods can not handle
non-stationarities. 
\end{myitemize}%
\myitem\cmt{summary}To sum up, none of the aforementioned schemes 
identifies interaction graphs in time-varying systems with memory. To
 the best of our knowledge, the only exception
 is~\cite{zaman2017onlinetopology}, but it can only cope with slowly changing
 VAR coefficients.
\end{myitemize}

\cmt{contribution}To alleviate these limitations, the present paper 
\begin{myitemize}
\myitem\cmt{estimator\ra interpretable graph + forecasting}relies on a 
 time-varying VAR (TVAR) model to propose a novel estimation criterion
 for non-stationary data that accounts for memory and a network
 structure in the interactions. The resulting estimates provide
 allow forcasting and  \emph{impulse response causality} analysis~\cite[Ch.~2]{lutkepohl2005}. 
\myitem\cmt{local structural breakpoint}A major novelty is the  notion
 of \emph{local structural breakpoint}, which captures the intuitive fact that
 changes in the interactions  need not be synchronized
 across the system; in contrast to most existing works. 
\myitem\cmt{solver}Furthermore, a low-complexity solver is proposed to minimize the
 aforementioned criterion and a 
\myitem\cmt{windowing}windowing technique is proposed to accommodate
prior information on the system dynamics and reduce computational
complexity. 

\end{myitemize}

%\cmt{notation}
%Regarding \emph{ notation}, \begin{myitemize}
%\myitem Vectors are expressed as bold lowercase letters,
%\myitem matrices as bold uppercase letters.
%\myitem the notation $[m,n]$ with $m$ and $n$
%integers satisfying $m\leq n$ will stand for $\{m,m+1,\ldots,n\}$.
%\end{myitemize}
The rest of the paper is structured as follows. Sec.~\ref{sec:networkid} introduces the model and the proposed criterion, with some practical considerations in Secs.~\ref{sec:windowing} and~\ref{ss:choiceOfParameters}; and Sec.~\ref{sec:solver} presents an iterative solver. Numerical experiments are described in Sec.~\ref{sec:experiments} and conclusions in Sec.~\ref{sec:conclusion}.

\newcommand{\vy}[1]{\mathbf{y}_{#1}}
\newcommand{\mphi}[2]{\mathbf{A}^{(#1)}_{#2}}
\newcommand{\mphiw}[2]{\tilde{\mathbf{A}}^{(#1)}_{#2}}
\newcommand{\mpsi}[2]{\boldsymbol{\psi}^{(#1)}_{#2}}
\newcommand{\bPsi}{\boldsymbol{\Psi}}
\newcommand{\vzero}{\mathbf{0}}
\newcommand{\vyT}[1]{\vy{#1}^\top}
\newcommand{\mtheta}[1]{\boldsymbol{\theta}_{#1}}
\newcommand{\mthetaT}[1]{\mtheta{#1}^\top}
\newcommand{\veps}[1]{\boldsymbol{\varepsilon}_{#1}}
\newcommand{\vepsT}[1]{\veps{#1}^\top}
\newcommand{\vtepsT}[1]{{ \tilde{\boldsymbol{\varepsilon}}_{#1}^\top }}

\newcommand{\mSqrtSigma}[1]{\boldsymbol{\Sigma}_{#1}^{1/2}}
\newcommand{\vx}[1]{{\mathbf{x}_{#1}}}
\newcommand{\vxT}[1]{{\mathbf{x}^{\top}_{#1}}}
\newcommand\norm[1]{\left\lVert#1\right\rVert}
\newcommand{\glPenalty}{\Omega_{GL}}
\newcommand\eye{\mathbf{I}}
\newcommand{\prox}{\mathrm{prox}}
\newcommand{\remove}[1]{}
\newcommand{\bigRmatrix}{\begin{bmatrix}
    \eye   & \mathbf{R}_{q+1} & \mathbf{R}_{q+1,3} & \ldots & \mathbf{R}_{q+1,T} \\
    \vzero & \eye            & \mathbf{R}_{q+2} & \ldots & \mathbf{R}_{q+2,T} \\
    \vzero & \vzero          & \eye            & \ddots &   \vdots \\
    \vdots & \ldots          & \ddots          & \ddots &  \mathbf{R}_{T-1} \\
    \vzero &  \ldots         & \vzero          & \vzero  &\eye
    \end{bmatrix}}

\section{Dynamic network identification}
\label{sec:networkid}

After reviewing TVAR models and introducing the notion of
time-varying causality graphs, this section proposes an estimation
criterion and an iterative solver. Extensions and general
considerations are provided subsequently.

\subsection{Time-varying interaction graphs}
\label{sec:interactiongraphs}
\cmt{TVAR}
\begin{myitemize}
\myitem A multivariate  time series is a collection
$\{\vy{t}\}_{t=1}^T$ of vectors $\vy{t} := [y_{1, t}, y_{2,t}, \ldots,
y_{P, t}]^\top$. The $i$-th
(scalar) time series comprises the samples $\{y_{i,t}\}_{t=1}^T$ and
can correspond e.g. with the activity over time of the
$i$-th \emph{region of interest} in a brain network, or with the
measurements of the $i$-th sensor in a sensor network.
\myitem A customary model for multivariate time series generated by
non-stationary dynamic systems is the so-called $L$-th order
TVAR model \cite[Ch. 1]{lutkepohl2005}:\vspace{-2mm}
\begin{equation}\label{eq:varModel}
    \vy{t} = \sum_{\ell = 1}^{L} \mphi{\ell}{t} \vy{t - \ell}
    + \veps{t}\vspace{-2mm}
\end{equation}
where the matrix entries $\{a_{ij,t}^{(\ell)}\}_{i,j \in [1, P], t \in
[1, T]}$ are the model coefficients and $\veps{i,t}$ form the
innovation process. 
\cmt{def. of [1 P] can be moved to notation paragraph}
Throughout, the notation $[m,n]$ with $m$ and $n$
integers satisfying $m\leq n$ will stand for $\{m,m+1,\ldots,n\}$.
\myitem\cmt{time-invariant VAR}A time-invariant VAR model is a special
case of \eqref{eq:varModel} where $a_{ij,t}^{(\ell)}=
a_{ij,t^\prime}^{(\ell)}~\forall(t, t^\prime)$.
\end{myitemize}

\cmt{Time-varying filters}
\begin{myitemize}
\myitem\cmt{scalar form}An insightful interpretation of time-varying
VAR models stems from expressing  \eqref{eq:varModel} as
\begin{subequations}
\begin{align}
\label{eq:varScalarForm}
&    y_{i,t} = \textstyle\sum_{\ell = 1}^{L} \textstyle\sum_{j=1}^P a_{ij,t}^{(\ell)}
    y_{j,t-\ell} + \veps{i,t}\\
    &= \sum_{j=1}^P \left[y_{j,t-1}, y_{j, t-2}, \ldots, y_{j,t-L}\right]\,\mathbf{a}_{ij,t} + \veps{i,t}
\end{align}
\end{subequations}
where  $\mathbf{a}_{ij,t}:=[a_{ij,t}^{(1)}, a_{ij,t}^{(2)}, \ldots, a_{ij,t}^{(L)}]^\top$.
\myitem From~\eqref{eq:varScalarForm}, the $i$-th sequence
$\{y_{i,t}\}_{t = 1}^T$ equals the innovation plus the sum of all
sequences $\{\{y_{p,t}\}_{t=1}^T\}_{p=1}^P$ after being filtered with
a \emph{linear time-varying} (LTV) filter with coefficients $\{a_{ij,t}^{(l)}\}_{l=1}^L$.
\end{myitemize}
        
\begin{figure}
    \centering \vspace{-6mm}\includegraphics[width=0.6\columnwidth]{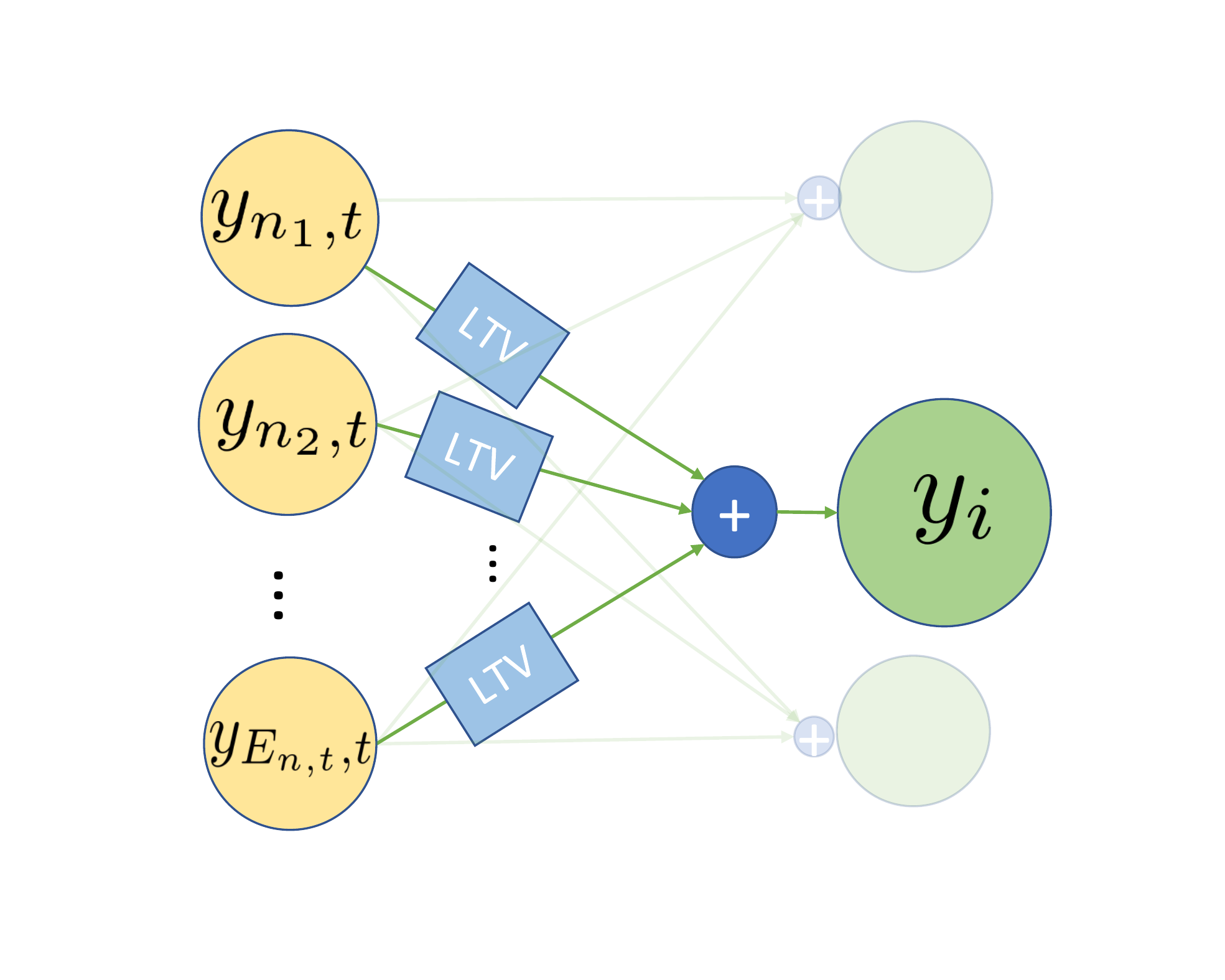}\vspace{-4mm} \caption{Graph associated with a TVAR model. %% \emph{Direct} influences over node $i$ from its neighboring nodes $n_1, \ldots, E_{n,t}$ are highlighted.\vspace{-3mm}
    }
    \label{fig:graphSketch}
\end{figure}

\cmt{Time-varying graph}
\begin{myitemize}
\myitem\cmt{motivation}As described in Sec.~\ref{sec:intro},
interactions among time series are generally indirect, which
translates into many of these LTV filters being
identically zero. To mathematically capture this interaction pattern, previous works consider the notion of graph associated
with a time-invariant VAR process (see
e.g.~\cite{bolstad2011groupsparse}), which is generalized next
to \emph{time-varying} VAR models~\eqref{eq:varModel}.
\myitem\cmt{definition}To this end, identify the $i$-th time series
with the $i$-th vertex (or node) in the vertex set $\mathcal{V}:=[1,
P]$ and define the time-varying edge set as
$\mathcal{E}_t:=\{(i,j)\in \mathcal{V} \times \mathcal{V}:\mathbf{a}_{ij,t} \neq \vzero\}$.
Thus, each edge of this time-varying graph can be thought of as an LTV
filter, as depicted in
Fig. \ref{fig:graphSketch}.
%\acom{impulse-response causality}

%% \myitem If the influences from node $i$ into node $j$ in the model above is originated by a network connection (link, edge), it makes sense to say that $\mathbf{a}_{ij,t}$ is the filter (impulse response) associated with the edge $(i,j)$. 
%% \myitem If this is the case, the \emph{(time-varying) graph} $\mathcal{G}$ associated with (underlying) the model is defined as the tuple $\mathcal{G}= \{\mathcal{V}:=[1, P], \mathcal{E}_t:=\{(i,j):\mathbf{a}_{ij,t} \neq \vzero\} \}$. 
\end{myitemize}

%% \begin{myitemize}

%% \myitem\begin{myitemize}
%% \myitem This implies that if $(i,j) \notin  \mathcal{E}_t$, then $\mathbf{a}_{ij,t} = \vzero$.
%% %if there is not an edge from $i$ to $j$, then $\mathbf{a}_{ij,t} = \vzero$.
%% \myitem This implies that the nonzero pattern of the VAR model matrices in \eqref{eq:varModel} verifies $\mathrm{supp}(\mphi{\ell}{t}) \subset \mathcal{E}_t \forall \ell, t$.
%% \end{myitemize}
%% \myitem This form is illustrated in Fig. \ref{fig:graphSketch}.
%% \end{myitemize}

%Let $\{\{y_i[t]\}_{i=1}^P\}_{t=1}^T$ be a sample of $T$ observations of a set of $p$ time series. For every time instant $t$, define $\vy{t} $ as the $p$-vector observed value of the VAR process at time $t$, and let $\mphi{\ell}{t}$ be the (time-varying) coefficient matrix corresponding to the $\ell$-th lag during the VAR segment where $t$ is located. A $q$-th order VAR process is modeled as 

%where $\veps{t}$ denotes multivariate Gaussian white (i.i.d.) noise, and $\mSqrtSigma{t}$ is the noise covariance matrix, both at time $t$.

%\clearpage
\subsection{Proposed estimation criterion}
\cmt{overview}
\begin{myitemize}
\myitem\cmt{goal}The main goal of this paper is to
estimate $\{\{\mphi{\ell}{t}\}_{\ell=1}^L\}_{t=L+1}^T$ 
%\acom{chec klimits} 
% Luismi: LIMITS ARE OK
given
$\{\vy{t}\}_{t=1}^T$.
\myitem\cmt{raw problem underdetermined}Without 
    additional assumptions, reasonable estimates cannot be found
    because the number of unknowns is $(T-L)P^2L$ whereas the number of
    samples is just $PL$.
\myitem\cmt{estimator\ra based on}This difficulty is typically
    alleviated by assuming certain structure usually found in
    real-world dynamic systems. As detailed next, the structure
    adopted here embodies 
\begin{myitemize}
\myitem\cmt{spatial structure}both the  sparsity of causal interactions
\myitem\cmt{temporal structure}and the spatial locality of changes in
    those interactions. 
\end{myitemize}%
\end{myitemize}%

\cmt{criterion}
The proposed estimation criterion is given by\vspace{-2mm}
\begin{align}\label{eq:criterionScalar}
    %\{\mathbf{A}^{\star(\ell)}_{t}\}_{\ell = 1, t =1}^{L, T} := \arg
    \min_{\{\mphi{\ell}{t}\}} &
    \sum_{t=L+1}^T \norm{\vy{t} - \sum_{\ell = 1}^{L} \mphi{\ell}{t} \vy{t - \ell}}_2^2 
    \\ \nonumber + \sum_{(i,j)} &\bigg(
      \lambda \sum_{t=L+1}^T \norm{\mathbf{a}_{ij,t}}_2 
    + \gamma  \sum_{t=L+2}^T \norm{\mathbf{a}_{ij,t}-\mathbf{a}_{ij,t-1}}_2\bigg) %\\
    %\mathrm{s.to}\;\; & [\mphi{(\ell)}{t}]_{ij} = [\mathbf{a}_{ij,t}]_{\ell} \; \; \forall (i, j, t, \ell)
    \vspace{-2mm}
\end{align}
where the first term promotes estimates that fit the data and the two
regularizers in parentheses are explained next. The regularization parameters $\lambda>0$ and $\gamma>0$ can be selected through
cross-validation to balance the relative weight of data and prior
information.

\begin{myitemize}
\myitem\cmt{spatial structure}The first regularizer is a group-lasso
penalty that promotes edge sparsity or, equivalently, that a large
number of LTV filters $\mathbf{a}_{ij,t}$ are $\bm 0$.  As delineated
in Secs.~\ref{sec:intro} and~\ref{sec:interactiongraphs}, this
corresponds to the intuitive notion that most interactions in a
complex network are indirect and therefore nodes are connected only
with a small fraction of other nodes. This regularizer generalizes the
one in~\cite{bolstad2011groupsparse} to time-varying graphs. 

\myitem\cmt{temporal structure}The second regularizer
\begin{myitemize}
\myitem\cmt{breakpoints}is a total-variation regularizer
that promotes estimates where the LTV filters remain constant over
time except for a relatively small number of time instants
$\mathcal{T}_{i,j}:=\{t: a_{ij,t}^{(\ell)} \neq a_{ij,t-1}^{(\ell)}$
for some $\ell \}$ denoted as \emph{local breakpoints}. This regularizer,
together with the notion of local breakpoints, constitutes one of the
major novelties of this paper and contrasts with the notion of
structural (or global) breakpoints, defined as
$\mathcal{T}:=\{t: \mphi{\ell}{t} \neq
\mphi{\ell}{t-1}$for some $\ell \}$ and adopted in the literature; see e.g.~\cite{safikhani2017structuralbreak,
tank2017efficient,ombao2005slex,aue2009break,cho2016change}. These works promote solutions
with few global breakpoints, and therefore all the LTV filter
estimates change simultaneously at the same time for all nodes. In
contrast, this work advocates promoting solutions with
few \emph{local} breakpoints, since it is expected that changes in
the underlying dynamic system take place locally. For instance, in a chemical process,
closing a valve between tank A and B affects the future interactions between
their pressures, but does not generally affect interactions between
the pressure of tanks C and D.

%% \cmt{Local breakpoints} A \emph{local breakpoint} will be defined here as the tuple $(i, j, t)$ such that $a_{ij,t}^{(\ell)} \neq a_{ij,t-1}^{(\ell)}$ for some $\ell \in [1, L]$. To distinguish them from the breakpoints defined in \cite{safikhani2017structuralbreak, tank2017efficient}, the latter will be referred to as \emph{global breakpoints}. Although a local breakpoint localized at $(i,j)$ means a local change in the coefficients, this is translated into a change in the interactions between other pairs of nodes whose influence is mediated by the link $(i,j)$. In other words, if a change in the global behavior of a network can be explained by a local breakpoint, this is simpler (and thus preferable from an informational point of view) than a global breakpoint involving more coefficient changes.
%% \myitem\cmt{regularizer}The second regularizer is a group total-variation (GTV) penalty that promotes solutions with a small number of local breakpoints, i.e. allows only a small amount of filters to vary across time instant $t$. Related penalties have been used to recover piecewise-constant signals in related problems such as denoising.
\end{myitemize}
\end{myitemize}

\subsection{Data windowing}
\label{sec:windowing}
\cmt{Overview}In practice, the time series are expected to evolve 
at a faster time scale than the underlying system that generates
them. In many applications, such as control of industrial processes,
the opposite would imply that the sampling interval needs to be
increased. If this is the case, it may be beneficial to assume that
$\mphi{\ell}{t}$ remain constant within a certain window since this
would decrease the number of coefficients to estimate and therefore
would improve estimation performance. 

\newcommand{\windowset}{\hc{\mathcal{W}}}
\newcommand{\windowind}{{\hc{{n}}}}
\newcommand{\windownum}{{\hc{{N}}}}
\newcommand{\cvnum}{{\hc{{M}}}} % crossvalidation number
\newcommand{\cvind}{{\hc{{m}}}} % crossvalidation number

\cmt{technique description}To introduce this windowing technique
let $\windowset_\windowind \subset [L+1,T]$ denote the set of indices in
the $\windowind$-th window, $\windowind=1,\ldots,\windownum$, and let
$\windowind(t)=\windowind_0$ if $t\in \windowset_{\windowind_0}$. If
$\mphi{\ell}{t}=\mphiw{\ell}{\windowind(t)}~\forall t$,
then \eqref{eq:criterionScalar} becomes
\begin{align}\label{eq:criterionScalarwindowed}
    %\{\mathbf{A}^{\star(\ell)}_{t}\}_{\ell = 1, t =1}^{L, T} := \arg
&     \min_{\{\mphiw{\ell}{t}\}} 
    \textstyle\sum_{t=L+1}^T \norm{\vy{t} - \textstyle\sum_{\ell = 1}^{L} \mphiw{\ell}{\windowind(t)} \vy{t - \ell}}_2^2 
 + \sum_{(i,j)} \\ \nonumber& \bigg(
      \lambda \sum_{t=L+1}^T \norm{\tilde{\mathbf{a}}_{ij,\windowind(t)}}_2 
       + \gamma  \sum_{t=L+2}^T \norm{\tilde{\mathbf{a}}_{ij,\windowind(t)}-\tilde{\mathbf{a}}_{ij,\windowind(t-1)}}_2\bigg) %\\
    %\mathrm{s.to}\;\; & [\mphiw{(\ell)}{t}]_{ij} = [\mathbf{a}_{ij,t}]_{\ell} \; \; \forall (i, j, t, \ell)
\end{align}
where $\tilde{\mathbf{a}}_{ij,t}$ is correspondingly defined in terms
of $\mphiw{\ell}{t}$. Absorbing scaling factors in the regularization
parameters, \eqref{eq:criterionScalarwindowed} boils down to 
\begin{align}\label{eq:criterionScalarwindowedsimplified}
    %\{\mathbf{A}^{\star(\ell)}_{t}\}_{\ell = 1, t =1}^{L, T} := \arg
     \min_{\{\mphiw{\ell}{t}\}} &
    \sum_{\windowind=1}^\windownum \sum_{t\in\windowset_\windowind}\norm{\vy{t} - \sum_{\ell = 1}^{L} \mphiw{\ell}{\windowind} \vy{t - \ell}}_2^2 
    \\ \nonumber + \sum_{(i,j)} &\bigg(
      \lambda     \sum_{\windowind=1}^\windownum \norm{\tilde{\mathbf{a}}_{ij,\windowind}}_2 
    + \gamma     \sum_{\windowind=2}^\windownum \norm{\tilde{\mathbf{a}}_{ij,\windowind}-\tilde{\mathbf{a}}_{ij,\windowind-1}}_2\bigg). %\\
    %\mathrm{s.to}\;\; & [\mphiw{(\ell)}{t}]_{ij} = [\mathbf{a}_{ij,t}]_{\ell} \; \; \forall (i, j, t, \ell)
\end{align}
Note that, while $LP^2(T-L)$ coefficients need to
be estimated in \eqref{eq:criterionScalar}, this number reduces
to  $LP^2N$
in \eqref{eq:criterionScalarwindowedsimplified}.

\cmt{discussion}Besides an improvement in the estimation
performance~\eqref{eq:criterionScalarwindowedsimplified} when the
length of the windows is attuned to the dynamics of the system, it can
be shown that the objective function becomes strongly convex if
windows are sufficiently large, which speeds up the convergence of the
algorithm in Sec.~\ref{sec:solver} (convergence becomes linear). The
caveat here is a loss of temporal resolution: if one wishes to detect
local breakpoints and two or more changes are produced in the same LTV
filter within a single window, then the algorithm will only detect at most a
single breakpoint. This effect can be counteracted by
applying a screening techniques along the lines of \cite{safikhani2017structuralbreak}.

\begin{figure*}[!tb]
    \centering
    \includegraphics[width=\textwidth]{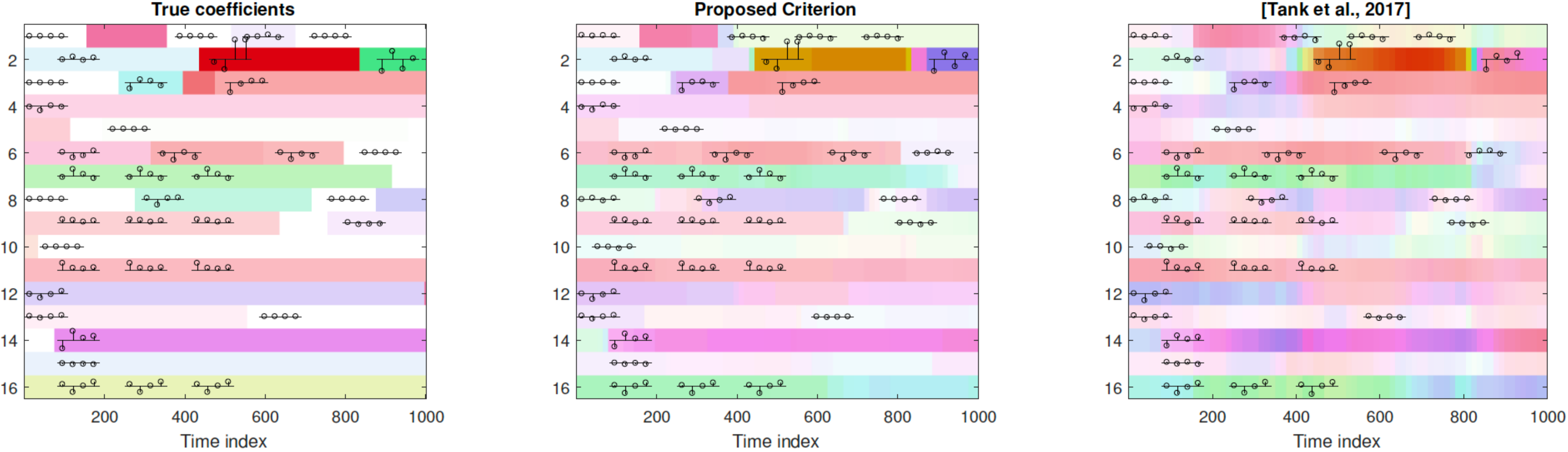}
    \caption{
 Comparison between the estimates of the proposed criterion and the one in (Tank et al., 2017) \cite{tank2017efficient}}\vspace{-5mm}
    \label{fig:proposedVsTank}
\end{figure*}

\subsection{Choice of parameters}
\label{ss:choiceOfParameters}

\cmt{overview}Regularization parameters, in this case $\lambda$ and
$\gamma$, are conventionally set through cross-validation. However, such a task may be challenging when dealing with non-stationary data. If one decides to carry
out $\cvnum$-fold cross validation, forming $\cvnum$ sets of
consecutive samples is not appealing since the estimate of the
fitting error in the validation set will become artificially high
and not informative about  whether the algorithm is
learning changes in the VAR coefficients.

\cmt{proposed CV technique}To circumvent this limitation, the proposed
technique forms the aforementioned sets by skipping one out of
$\cvnum$ time samples. The estimator for the $\cvind$-th fold becomes
%\acom{pls make indices of second sum smaller--DONE}
\begin{align}\label{eq:criterionScalarwindowedsimplifiedcv}
    %\{\mathbf{A}^{\star(\ell)}_{t}\}_{\ell = 1, t =1}^{L, T} := \arg
     \min_{\{\mphiw{\ell}{t}\}} &
    \sum_{\windowind=1}^\windownum \underset{t~\mathrm{mod} \cvnum \neq \cvind}{\sum_{t\in\windowset_\windowind}}
   % \sum_{\begin{array}{c}\small{ t\in\windowset_\windowind}\\
     %t~\mathrm{mod} \cvnum \neq \cvind \end{array}}
     \norm{\vy{t}
     - \sum_{\ell = 1}^{L} \mphiw{\ell}{\windowind} \vy{t - \ell}}_2^2 
    \\ \nonumber + \sum_{(i,j)} &\bigg(
      \lambda     \sum_{\windowind=1}^\windownum \norm{\tilde{\mathbf{a}}_{ij,\windowind}}_2 
    + \gamma     \sum_{\windowind=2}^\windownum \norm{\tilde{\mathbf{a}}_{ij,\windowind}-\tilde{\mathbf{a}}_{ij,\windowind-1}}_2\bigg). %\\
    %\mathrm{s.to}\;\; & [\mphiw{(\ell)}{t}]_{ij} = [\mathbf{a}_{ij,t}]_{\ell} \; \; \forall (i, j, t, \ell)
\end{align}
Admittedly, all vectors $\{\vy{t}\}_t$ will still show up in all
folds, but only as regressors in those folds where they are not target vectors. Indeed, this
does not cause any problem from a theoretical standpoint and the
performance observed in the numerical tests supports this approach.

%% \begin{myitemize}
%% \myitem The regularization parameters $\lambda$ and $\gamma$ will regulate the number of breakpoints and the total amount of edges with nonzero coefficients.
%% \myitem We want our approach to be purely-data based.
%% \myitem The choice of the hyperparameters will be done via cross-validation.
%% \myitem The standard N-fold cross-validation consists in dividing the available data in $N$ subsets, and calculating the average fitting error over the $N$ folds
%% \myitem The $n$th \emph{fold} consists in estimating the model parameters with all data subsets except the $n$th, and computing the fitting error on the $n$ data subset.
%% \myitem Since the data are dependent and the model order is $L$, every data point that is excluded from a training set subtracts $L+1$ training points.
%% \myitem To ensure that the cross-validation is proper, every data point $\vy{t}$ must not be used as a training \emph{input and output} inside the same fold.
%% \myitem This is ensured if every fold is assigned a subset of rows of the matrices in \eqref{eq:criterionMatrix}. In particular, the $i$th row is deleted from the matrices in the $n$-th fold iff $1 = i (\mathrm{ mod }\; n)$.
%% \end{myitemize}

\subsection{Iterative solver}
\label{sec:solver}
%\acom{polish this section}

This section outlines the derivation of an ADMM-based algorithm proposed to
solve~\eqref{eq:criterionScalar}. Define
\cmt{matrix form} \begin{myitemize}
\myitem $\mathbf{Z}$ as the block-diagonal matrix with blocks $(\vxT{q+1},$ $ \vxT{q+2}, \ldots, \vxT{T})$, with 
$\vxT{t}:= [\vyT{t-1} \ldots \vyT{t-q}]$; 
\myitem $\mathbf{B} := [\mathbf{B}_{q+1}^\top, \ldots,$ $ \mathbf{B}_{T}^\top]$, with $ 
    \mathbf{B}_{t} := \big[
    \mphi{1}{t}, 
    \mphi{2}{t}, \ldots ,
    \mphi{q}{t}
    \big]^\top
$; and $\mathbf{Y}:= \left[\vy{q+1},\ldots, \vy{T}\right]^\top$.
\myitem Then \eqref{eq:criterionScalar} can be rewritten as
\begin{equation}\label{eq:criterionMatrix}
    \arg\min_{\mathbf{B}} \frac{1}{2} \norm{\mathbf{Y}-\mathbf{Z} \mathbf{B}}_{F}^2 + \lambda \glPenalty(\mathbf{B}) + \gamma \Omega_{GTV}(\mathbf{B})
\end{equation}
\myitem where $\glPenalty(\mathbf{B}) = \sum_{(i,j)} \sum_{t=L+1}^T \norm{\mathbf{a}_{ij,t}}_2$, 
\myitem and   $\Omega_{GTV}(\mathbf{B}) = \sum_{(i,j)} \sum_{t=L+1}^T \norm{\mathbf{a}_{ij,t+1}-\mathbf{a}_{ij,t}}_2$.
\myitem Upon defining
\renewcommand{\arraystretch}{0.5}
\begin{align*}
%\scriptstyle
    \mathbf{D} := \begin{bmatrix}
    -\eye  & \eye   & \vzero & \ldots & \vzero \\
    \vzero & -\eye  & \eye   &        & \vdots \\
    \vdots & \ddots & \ddots & \ddots &        \\
    \vzero &        & \ldots & -\eye  & \eye   
    \end{bmatrix}, \vspace{2mm}
\end{align*}
$\Omega_{GTV}(\mathbf{B})$ can be expressed as $\glPenalty(\mathbf{D}\mathbf{B})$.
%\begin{equation}
%    \arg\min_{\mathbf{B}} \frac{1}{2} \norm{\mathbf{Y}-\mathbf{Z} \mathbf{B}}_{F}^2 + \lambda \Omega_1(\mathbf{B}) + \gamma \Omega_1(\mathbf{D} \mathbf{B})
%\end{equation}
\end{myitemize}
This allows to rewrite \eqref{eq:criterionMatrix} along the lines of \cite{wahlberg2012admm} for solving via ADMM
\begin{align}
    \arg\min_{\mathbf{B}, \mathbf{\Theta}, \mathbf{C}} &\; \frac{1}{2} \norm{\mathbf{Y}-\mathbf{Z} \mathbf{B}}^2_{F} + \lambda \nonumber \glPenalty(\mathbf{\Theta}) + \gamma \glPenalty(\mathbf{C}), \\
    %\nonumber 
    \mathrm{s. to} &\;  \mathbf{D} \mathbf{B} = \mathbf{\Theta}, %\\
    %& 
    \; \mathbf{B} = \mathbf{C}
\end{align}
%\setcounter{MaxMatrixCols}{20}
%\renewcommand*{\arraystretch}{1.5}
%The augmented Lagrangian is $\mathscr{L}_{\rho} (\mathbf{B}, \mathbf{\Theta}, \mathbf{C}; \mathbf{U}, \mathbf{V}) $
%\begin{align}
%    :=  \frac{1}{2} \norm{\mathbf{Y}-\mathbf{Z} \mathbf{B}}_{F}^2 & + \lambda \glPenalty(\mathbf{\Theta}) + \gamma \glPenalty (\mathbf{C}) \\  + \rho \; \mathrm{tr}(\mathbf{U}^\top (\mathbf{B}-\mathbf{C}) &+ \mathbf{V}^\top (\mathbf{D}\mathbf{B}-\mathbf{\Theta})) 
%    \\  + \frac{\rho}{2} (\norm{\mathbf{B}-\mathbf{C}}_F^2 &+ \norm{\mathbf{D} \mathbf{B}-\mathbf{\Theta}}_F^2)
%\end{align}

The ADMM for the $\rho$-augmented Lagrangian with scaled dual variables $\mathbf{U}$ and $\mathbf{V}$ computes for each iteration $k$:
\begin{align}
\label{eq:update_B}
    \mathbf{B}^{[k+1]} & := %\arg\min_{\mathbf{B}} \frac{1}{2} \norm{\mathbf{Y}-\mathbf{Z} \mathbf{B}}_{F}^2 + \frac{\rho}{2} \norm{\mathbf{B}-\mathbf{C}^{[k]} + \mathbf{V}^{[k]}}_F^2 + \frac{\rho}{2} \norm{\mathbf{D}\mathbf{B}-\mathbf{\Theta}^{[k]} + \mathbf{U}^{[k]}}_F^2 \\ 
    %\nonumber & = \left( \mathbf{Z}^\top \mathbf{Z} + \rho (\eye + \mathbf{D}^\top \mathbf{D})\right) ^\dagger \left( \mathbf{Z}^\top \mathbf{Y} + \rho (\mathbf{C}^{[k]} - \mathbf{V}^{[k]} + \mathbf{D}^\top(\mathbf{\Theta}^{[k]} - \mathbf{U}^{[k]}))\right) \\
    \left( \mathbf{Z}^\top \mathbf{Z}/\rho + \eye + \mathbf{D}^\top \mathbf{D}\right) ^\dagger \\ &\nonumber \big( \mathbf{Z}^\top \mathbf{Y}/\rho + \mathbf{C}^{[k]} - \mathbf{V}^{[k]} + \mathbf{D}^\top(\mathbf{\Theta}^{[k]} - \mathbf{U}^{[k]})\big)\\
\boldsymbol{\theta}_{ij,t}^{[k+1]} &:= \prox_{\lambda/\rho\norm{\cdot}_2}({\mathbf{b}_{ij,t}^{[k+1]} - \mathbf{b}_{ij,t-1}^{[k+1]} -     \mathbf{u}_{ij,t-1}^{[k+1]} }) \label{eq:computeThetat} \\
\mathbf{c}_{ij,t}^{[k+1]} &:= \prox_{\lambda/\rho\norm{\cdot}_2}({\mathbf{b}_{ij,t}^{[k+1]} - \mathbf{v}_{ij,t}^{[k+1]} })\label{eq:computeCt}\\
%% \label{eq:update_Theta}
%%     \mathbf{\Theta}^{[k+1]} & := %\arg\min_{\mathbf{\Theta}} \lambda \glPenalty(\mathbf{\Theta}) + \frac{\rho}{2} \norm{\mathbf{D}\mathbf{B}^{[k+1]}-\mathbf{\Theta} + \mathbf{U}^{[k]}}_F^2 \\
%%     %\nonumber&= 
%%     \prox_{\lambda/\rho\glPenalty}(\mathbf{D}\mathbf{B}^{[k+1]} - \mathbf{U}^{[k]})\\
%% \label{eq:update_C}
%%     \mathbf{C}^{[k+1]} & := %\arg\min_{\mathbf{C}} \gamma \glPenalty(\mathbf{C}) +
%%     %\frac{\rho}{2} \norm{\mathbf{B}^{[k+1]}-\mathbf{C} + \mathbf{V}^{[k]}}_F^2\\ 
%%     %\nonumber&= 
%%     \prox_{\gamma/\rho\glPenalty}(\mathbf{B}^{[k+1]} - \mathbf{V}^{[k]}) \\
    \mathbf{U}^{[k+1]} &:= \mathbf{U}^{[k]} + (\mathbf{D}\mathbf{B}^{[k+1]} - \mathbf{\Theta}^{[k+1]}) \label{eq:updateU}\\
    \mathbf{V}^{[k+1]} &:= \mathbf{V}^{[k]} + (\mathbf{B}^{[k+1]} - \mathbf{C}^{[k+1]})\label{eq:updateV}
\end{align}
and it is summarized in Proc. \ref{alg:admm}. The
update \eqref{eq:update_B} can be efficiently computed by exploiting
the tri-diagonal strucutre of $\mathbf{Z}$ and $\mathbf{D}$.  The
updates in \eqref{eq:computeThetat} and \eqref{eq:computeCt} exploit
the fact that the resulting prox operators are separable per $(i,j)$
and can be expressed in terms of a group-soft-thresholding
operator \cite{boyd2011distributed}.
%% \begin{align}
%% \end{align}

\remove{ 
The problem structure ($\mathbf{Z}^\top\mathbf{Z}$ is block diagonal, and $\mathbf{D}^\top\mathbf{D}$ is block-tridiagonal with $-\eye$ in the block-subdiagonals) %Thus, the evaluation of \eqref{eq:update_B} comes as the solution to a block-tridiagonal system and can be performed efficiently via %a banded linear system solver, or even faster by 
%an ad-hoc algorithm based on Thomas' tridiagonal solver.
allows to solve \eqref{eq:update_B} efficiently by first precomputing the matrices
\begin{equation}\label{eq:precomputeR}
\mathbf{R}_{t} := 
     (\mathbf{x}_{t} \vxT{t}/ \rho + 3\eye - \mathbf{R}_{t+1})^\dagger
\end{equation}
for $t = T, T-1, \ldots, q+2$, with $\mathbf{R}_T:=\eye$ and 
$\mathbf{R}_{1}:= 
    (\mathbf{x}_{q+1} \vxT{q+1}/ \rho + 2\eye - \mathbf{R}_{q+2})^\dagger $; and computing for each $k$
%\subsection{Efficient block-tridiagonal solver}
%\acom{this section may be removed or combined with the previous one}
\begin{align}
%\mathbf{Q}_T = & \mathbf{x}_{T} \vyT{T}/\rho + \mathbf{C}_T^{[k]} - \mathbf{V}_T^{[k]} \\ &+ \mathbf{\Theta}_{T-1}^{[k]} -\mathbf{\Theta}_T^{[k]} - \mathbf{U}_{T-1}^{[k]} + \mathbf{U}_{T}^{[k]} \nonumber \\ 
%&\forall i \in [q+2, T-1], \nonumber \\
\mathbf{Q}_i^{[k]} = & \mathbf{R}_{i+1} \mathbf{Q}_{i+1}^{[k]} + \mathbf{x}_{i} \vyT{i}/\rho + \mathbf{C}_{i}^{[k]} - \mathbf{V}_{i}^{[k]} \nonumber\\
&+ \mathbf{\Theta}_{i-1}^{[k]} -\mathbf{\Theta}_i^{[k]} - \mathbf{U}_{i-1}^{[k]} + \mathbf{U}_{i}^{[k]}  \label{eq:computeQ}\\
%\mathbf{B}_{q+1} = & \mathbf{R}_{q+1}\bigg(\mathbf{R}_{q+2} \mathbf{Q}_{q+2} + \mathbf{x}_{q+1} \vyT{q+1}/\rho \\
%& \nonumber + \mathbf{C}_{q+1}^{[k]} - \mathbf{V}_{q+1}^{[k]} -\mathbf{\Theta}_{q+1}^{[k]} + \mathbf{U}_{{q+1}}^{[k]}\bigg) \\ 
%&\forall i \in [q+2, T], \\
\mathbf{B}_i^{[k]} = & \mathbf{R}_{i}(\mathbf{Q}_i^{[k]} + \mathbf{B}_{i-1}^{[k]}) \label{eq:computeBt}
\end{align}
is computed for each $k$ with $\mathbf{Q}_{T+1}^{[k]}:=0, \mathbf{B}_q^{[k]}=:0$, reducing the computational cost of \eqref{eq:update_B} to $8(T-L)$ additions and $2(T-L)$ multiplications of $pq\times pq$ matrices per ADMM iteration.
%%% ALGORITHMIC TABLE
%\acom{add commands in preamble}
%\usepackage{algorithm}
%\usepackage[noend]{algpseudocode}
%\floatname{algorithm}{Procedure}

\For{t \in [L+1, T]}
\State Update $\mathbf{c}_{ij,t}, \mathbf{\theta}_{ij,t-1}$ via (\ref{eq:computeThetat},\ref{eq:computeCt}) %\Comment{Group-soft thresholding}}
\EndFor
%\EndFor

}
\begin{algorithm}[t]
\caption{ADMM solver for dynamic network ID}\label{alg:admm}
\textbf{Input:} $\lambda, \gamma$, data $\{\vy{t}\}_{t =1}^{T}$\\
\textbf{Output:} matrix $\mathbf{B}$ containing VAR coefficients
\begin{algorithmic}[1]
\For{$k=1, \ldots$ until convergence}
\State Update $\mathbf{B}_t$ via \eqref{eq:update_B}\Comment{Block-Tridiagonal system}
\For{$t \in [L+1, T]$}
\For{$(i,j) \in [1,P]^2$} \Comment{Group-soft thresholding}
\State Update $\mathbf{c}_{ij,t}, \mathbf{\theta}_{ij,t-1}$ via (\ref{eq:computeThetat},\ref{eq:computeCt}) 
\EndFor
\EndFor
\State Update $\mathbf{U}, \mathbf{V}$ via (\ref{eq:updateU},\ref{eq:updateV}) \Comment{Dual update}
\EndFor
\end{algorithmic}
\end{algorithm}
%%%

\section{Numerical experiments}
\label{sec:experiments}

\newcommand\nNodes{\hc{4}}
\newcommand\erdosRenyiEdgeProbability{\hc{0.5}}
\newcommand\nObservations{\hc{1000}}
\newcommand\nBreakpoints{\hc{100}}
\newcommand\switchingProbability{\hc{0.4}}
\newcommand\filterOrder{\hc{4}}
\newcommand\syntheticSigma{\hc{0.03}}

A simple experiment is shown next to validate the proposed estimator.  An
Erdos-Renyi~\cite{kolaczyck2009} random graph $\mathcal{G}_0$ is
generated with $P=\nNodes$ nodes and an edge probability of
$P_0^{(i,j)} := \erdosRenyiEdgeProbability$ if $i\neq j$ and
$P_0^{(i,j)} := 0$ if $i=j$. 
%A randomly time-varying graph is generated from $\mathcal{G}_0$ as follows. First a subset $\mathcal{T}_b := \{t_{b1}, t_{b2}, \ldots, t{Nb}\} \in [q+1, T]$ where $T:=\nObservations$ is the length of the simulated time series, and  $N_b:=\nBreakpoints$ is the (fixed) number of breakpoints to be generated. Then, for each $t \in [q+1, T]$, if $t \in \mathcal{T}_b$, an node pair $(i,j)$ is selected at random and its edge is switched with probability $P_s:=\switchingProbability$; otherwise, $\mathcal{E}_t = \mathcal{E}_{t-1}$.
An $(L=\filterOrder)$-order TVAR model is generated, with initial VAR coefficients $\{\mathbf{A}^{(\ell)}_{L+1}\}_{\ell = 1}^{L}$ over $\mathcal{G}_0$ drawn from a standard normal distribution and scaled to ensure stability~\cite[chapter 1]{lutkepohl2005}.
Local breakpoints are generated at $N_b=
\nBreakpoints$ uniformly spaced time instants $\mathcal{T}_b :=
\{t_{b1}, t_{b2}, \ldots, t_{bN_b}\}$, and for each $t_b \in
\mathcal{T}_b$ a pair of nodes $(i_b, j_b)$ is selected uniformly at
random, generating a local breakpoint at the triplet $(t_b, i_b,
j_b)$.
For each breakpoint $b$, the VAR coefficients $\mathbf{a}_{i_b j_b,t_b}$ and the edge set $\mathcal{E}_t$ are changed as follows: if $(i_b, j_b) \in \mathcal{E}_{t_b-1}$, $\mathbf{a}_{i_b j_b,t_b}$ is set to $\vzero$ with probability $P_z:=\switchingProbability$; otherwise, a new standard Gaussian coefficient vector $\mathbf{a}_{i_b j_b,t_b}$ is generated and scaled to keep stability.
A realization of this TVAR process is generated by drawing $\{\vy{\ell}\}_{\ell=1}^L$ and $\{\veps{t}\}_{t=L+1}^T$ i.i.d from a zero-mean Gaussian distribution with variance  $\sigma_\epsilon^2:=\syntheticSigma$.

Fig.~\ref{fig:proposedVsTank} compares the true coefficients with the
estimates obtained by  the proposed criterion and the one in~\cite{tank2017efficient}. The latter only detects global (but not local) breakpoints. The windowing described in
Sec.~\ref{sec:windowing} selects subperiods of length $N=21$, and
$\lambda$ and $\gamma$ have been selected using the cross-validation
scheme described in Sec.~\ref{ss:choiceOfParameters}, both for the
proposed algorithm and the one in \cite{tank2017efficient} (which only
uses $\lambda$).

In each subfigure, each horizontal band corresponds to a pair of nodes, and the horizontal axis represents time. The LTV impulse response vectors $\mathbf{a}_{ij, t}/\norm{\mathbf{a}_{ij, t}}$ are mapped to colors in an HSV space, being assigned similar hue if their unitary counterparts $\mathbf{a}_{ij, t}/\norm{\mathbf{a}_{ij, t}}$ are closeby. The value (brightness) is set proportional to $\norm{\mathbf{a}_{ij, t}}$, so responses close to $\vzero$ appear close to white, whereas impulse responses with a larger $\ell_2$-norm will appear in a darker color. The stems appearing between some pairs of breakpoints represent filter coefficients of $\mathbf{a}_{ij, t}$ during the segment they lie on.

It is observed that the proposed algorithm could detect most of the
local breakpoints and correctly identifies segments of
stationarity. On the other hand, the competing algorithm yields a high
number of false positives as expected.

~\\[-1cm]
\section{Conclusions}
\label{sec:conclusion}
Dynamic networks can be identified using the notion of local
breakpoints, when VAR coefficient changes appear in a small number of
edges. The proposed technique involves three novelties: a regularized
criterion, a windowing technique, and a cross-validation
scheme. Simulation experiments encourage further research along these lines.

%% \section{To-do list}
%% \begin{itemize}
%% \item \acom{Should we use SPS template?}
%% \end{itemize}

\clearpage
%%%%%%%%%%%%% DO NOT MODIFY  %%%%%%%%%%%%%%%%%%%%%%%%%%%%%%
\if\editmode1 
\onecolumn
\printbibliography
\else
\bibliography{\bibfilenames}
\fi
\end{document}